\documentstyle [12pt,epsf] {article}

\newcommand{\sect}[1]{ \section{#1} \setcounter{equation}{0} }

\newcommand{\req}[1]{(\ref{#1})}
\newcommand{\nwc} {\newcommand}
\nwc{\hyp} {\hyphenation}
\nwc{\Z}{\ZZ}

\def\bfone{\relax{\rm 1\kern-.35em 1}}
\def\inbar{\vrule height1.5ex width.4pt depth0pt}
\def\IC{\relax\,\hbox{$\inbar\kern-.3em{\mss C}$}}
\def\ID{\relax{\rm I\kern-.18em D}}
\def\IF{\relax{\rm I\kern-.18em F}}
\def\IH{\relax{\rm I\kern-.18em H}}
\def\II{\relax{\rm I\kern-.17em I}}
\def\IN{\relax{\rm I\kern-.18em N}}
\def\IP{\relax{\rm I\kern-.18em P}}
\def\IQ{\relax\,\hbox{$\inbar\kern-.3em{\rm Q}$}}
\def\IR{\relax{\rm I\kern-.18em R}}
\def\ZZ{\relax{\hbox{\mss Z\kern-.42em Z}}}
\font\cmss=cmss10 \font\cmsss=cmss10 at 7pt
\def\ZZ{\relax\ifmmode\mathchoice
{\hbox{\cmss Z\kern-.4em Z}}{\hbox{\cmss Z\kern-.4em Z}}
{\lower.9pt\hbox{\cmsss Z\kern-.4em Z}}
{\lower1.2pt\hbox{\cmsss Z\kern-.4em Z}}\else{\cmss Z\kern-.4em Z}\fi}

\nwc{\be}  {\begin{equation}}
\nwc{\ee}  {\end{equation}}
\nwc{\ba}  {\begin{array}}
\nwc{\ea}  {\end{array}}
\nwc{\bdm} {\begin{displaymath}}
\nwc{\edm} {\end{displaymath}}
\nwc{\bea} {\be\ba{lcl}}
\nwc{\eea} {\ea\ee}
\nwc{\bda} {\bdm\ba{lcl}}
\nwc{\eda} {\ea\edm}
\nwc{\bc}  {\begin{center}}
\nwc{\ec}  {\end{center}}
\nwc{\ds}  {\displaystyle}
\nwc{\bmat}{\left(\ba}
\nwc{\emat}{\ea\right)}
\nwc{\nn} {\nonumber}
\nwc{\nnn} {\nonumber \vspace{.2cm} \\ }
\nwc{\ra}{\rightarrow}
\nwc{\lra}{\longrightarrow}
\nwc{\p} {\partial}

\nwc{\ep} {\epsilon}
\nwc{\de} {\delta}
\nwc{\Th} {\Theta}
\nwc{\th} {\theta}
\nwc{\al} {\alpha}
\nwc{\si} {\sigma}
\nwc{\Si} {\Sigma}
\nwc{\om} {\omega}
\nwc{\Om} {\Omega}
\nwc{\Ga} {\Gamma}
\nwc{\ga} {\gamma}
\nwc{\bet} {\beta}
\nwc{\La} {\Lambda}
\nwc{\la} {\lambda}
\nwc{\Sc}  {{\cal S}}
\nwc{\Rc}  {{\cal R}}
\nwc{\Dc}  {{\cal D}}
\nwc{\Oc}  {{\cal O}}
\nwc{\Cc}  {{\cal C}}
\nwc{\gc}  {{\cal g}}
\nwc{\Pc}  {{\cal P}}
\nwc{\Mc}  {{\cal M}}
\nwc{\Ec}  {{\cal E}}
\nwc{\Fc}  {{\cal F}}
\nwc{\Hc}  {{\cal H}}
\nwc{\Kc}  {{\cal K}}
\nwc{\Wc}  {{\cal W}}

\nwc{\Xc}  {{\cal X}}
\nwc{\Gc}  {{\cal G}}
\nwc{\Zc}  {{\cal Z}}
\nwc{\Nc}  {{\cal N}}

\nwc{\xc}  {{\cal x}}
\nwc{\Ac}  {{\cal A}}
\nwc{\Bc}  {{\cal B}}
\nwc{\Uc} {{\cal U}}
\nwc{\Vc} {{\cal V}}
\nwc{\Lc} {{\cal L}}
\nwc{\Qc} {{\cal Q}}

\nwc{\lng} {\langle}
\nwc{\rng} {\rangle}

\nwc{\lf} {\left}
\nwc{\ri} {\right}

\nwc{\diag} {{\rm diag}}
\nwc{\inv}  {{\rm inv}}
\nwc{\mod}  {{\ \rm mod\ }}
\nwc{\dete}  {{\rm det}}
\nwc{\tr}  {{\rm tr}}
\nwc{\im}  {{\rm Im}}
\nwc{\re}  {{\rm Re}}

\nwc{\h} {\frac{1}{2}}
\nwc{\fc} {\frac}

\def\KK{\relax{\rm I\kern-.18em K}}
\def\RR{\relax{\rm I\kern-.18em R}}
\def\NN{\relax{\rm I\kern-.18em N}}
\def\PP{\relax{\rm I\kern-.18em P}}

\def\zz{\relax{\sf Z\kern-.3em Z}}
\def\ZZ{\relax{\sf Z\kern-.4em Z}}
\def\ZZZ{{\relax{\sf Z}\kern -.5em Z}}
\def\ZZZ{Z\kern -0.37em Z}
\def\QQ{{\rm \kern .25em
             \vrule height1.4ex depth-.12ex width.06em\kern-.31em Q}}
\def\CC{{\rm \kern .25em
             \vrule height1.4ex depth-.12ex width.06em\kern-.31em C}}
\if@twoside\oddsidemargin25mm
\else\oddsidemargin25mm\evensidemargin25mm\marginparwidth25mm\fi%
\begin{document}
\begin{titlepage}
{\sf
\begin{flushright}
{CERN--TH/96--263}\\
{NEIP--96/007}\\
{hep--th/9610053}\\
{October 1996}
\end{flushright}}
\vfill
\vspace{-1cm}
\begin{center}
{\large \bf Self--Dual Strings and  Stability of BPS States\\[5mm]
in N=2\ SU(2)\ Gauge Theories$^{\mbox{\boldmath $\ast$}}$}
\vskip 1.2cm
{\sc A. Brandhuber$^{1,2}$} {\ \small \ and}
{\ \ \sc S. Stieberger$^{2}$}\\
\vskip 1.5cm
{\em $^1$CERN -- Theory Division} \\
{\em CH--1211 Gen\`eve 23, SWITZERLAND}
\vskip 1cm
{and}\\
\vskip .6cm
{\em $^2$Institut de Physique Th\'eorique}\\
{\em Universit\'e de Neuch\^atel}\\
{\em CH--2000 Neuch\^atel, SWITZERLAND}
\end{center}
\vfill

\thispagestyle{empty}

\begin{abstract}
We show how BPS states of supersymmetric $SU(2)$ Yang--Mills
with matter --both massless and massive--
are described as self--dual strings on a Riemann surface.
This connection enables us to prove the stability 
and the strong coupling behaviour of these states.
The Riemann surface naturally arises from type--IIB 
Calabi--Yau compactifications whose three--branes wrapped 
around vanishing two--cycles correspond 
to one--cycles on this surface.
\end{abstract}

\vskip 5mm \vskip0.5cm
\hrule width 5.cm \vskip 1.mm
{\small\small  $^\ast$ Supported by the
Swiss National Science Foundation and the EEC under contracts\\
SC1--CT92--0789 and European Comission TMR programme ERBFMRX--CT96--0045.\\
Email: Andreas.Brandhuber@cern.ch and stieberg@surya11.cern.ch}
\end{titlepage}

\sect{Introduction}

Within the last two years N=2 supersymmetric Yang--Mills theories
have received a lot of interest due to their ability of extracting non--trivial results about non--perturbative physics \cite{sw1,sw2,wl}.
In the following we will discuss $SU(2)$ supersymmetric Yang--Mills
with $N_f$ matter fields in the fundamental representation.
At the classical level, these theories contain the Abelian $U(1)$ 
gauge boson which gives rise to the N=2 vector multiplet $A$.
Besides we have the non--Abelian $W$--bosons together with 
$N_f$ quarks. The latter appear in N=2 hypermultiplets, which
can be arranged in two N=1 chiral multiplets 
$(Q^i,\tilde Q_i),\ \ i=1,\ldots,N_f$.
At the semi--classical level one can construct in the 
Coulomb phase monopole solutions and other dyonic states $(1,n_e)$, where
$n_e$ refers to their  electric charge $n_e$ and $n_m=1$ is the magnetic 
charge.
The construction of supersymmetric solutions for dyons with higher 
than one magnetic quantum number $(n_m=2)$ 
was presented for N=4 in \cite{sen} and for N=2 in \cite{zaslow}.
With an N=2 invariant mass term, the superpotential reads:

\be\label{superW}
W=\sqrt{2} A Q^i \tilde Q_i + \sum_{i=1}^{N_f} m_i Q^i  \tilde Q_i\ .
\ee
The mass of all states in such theories satisfy the Bogomolnyi--bound

\be\label{bpsmass}
M_{n_m,n_e,S_i} \geq \sqrt{2} |Z|=\sqrt{2} |n_m a_D(u)+n_e a(u)+
\fc{1}{\sqrt 2}\sum_iS_i m_i|\ ,
\ee
where $Z$ is the central charge of the N=2 supersymmetry algebra
which is a linear combination of conserved charges. The periods
$a(u)$ and $a_D(u)$ are holomorphic sections over the moduli
space $u$. The quantum numbers $S_i$ denote the global $U(1)$ charges
of the hypermultiplets. 
Particles for which the equality holds, are called  BPS saturated states.
This is the case for the small representations of the N=2 SUSY algebra
like hypermultiplets.
The quarks $(0,1)$ and the dyons $(1,n_e)$,\ $(2,n_e)$ appear in N=2 
hypermultiplets and are therefore BPS states with: $M=\sqrt{2} |Z|$.

In contrast to N=4 supersymmetric theories, these functions receive 
quantum corrections in N=2 theories, which fortunately are under control 
\cite{sw1,sw2,klt,bs,ito,bf2}.
Nevertheless, they have non--trivial dependence on the modulus $u$
and it may happen that the mass \req{bpsmass} of a dyon
becomes heavier than the sum of the mass of several dyons. Then it will decay
into these particles.
Of course, this decay has to conserve the quantum numbers $n_e,n_m,S_i$.
From eq. \req{bpsmass} it is easy to see this effect 
to take place for $m_i=0$  at the curve, determined by:

\be\label{curve}
\im\lf(\fc{a_D(u)}{a(u)}\ri)=0\ .
\ee
The fact that the BPS spectrum jumps when passing this curve, called
curve of marginal stability, is not expected in conventional field theories. 
This curve divides the moduli space into two parts: a strong coupling region 
and a weak coupling regime. States which are stable at weak coupling as e.g.
the gauge bosons decay when crossing this curve and we have to distinguish
between a weak--coupling spectrum and a strong--coupling spectrum
surving when passing this curve.
In general, at the classical level there are subspaces of the moduli
space, where parts or the full gauge group 
is restored and the $W$--bosons become massless. 
For $SU(2)$ gauge symmetry this is just the point $u=0$.
However, the non--perturbative expressions for $a_D(u)$ and $a(u)$ tell us that
there are points $u_j$ in the moduli space where hypermultiplets 
of spin $\leq \h$ and quantum numbers $(n^j_e,n^j_m)$
become massless. At these points we have:

\be\label{bound}
\lf.\lf(\fc{a_D(u)}{a(u)}\ri)\ri|_{u=u_j}=-\fc{n^j_e}{n^j_m}\ .
\ee
This means that these points lie on that curve.
Since it should be possible to go from $u=\infty$, the weak--coupling region,
until these points without crossing that curve, the states $(n^j_e,n^j_m)$
are supposed to exist also at weak--coupling.
Notify that, except at the superconformal points \cite{ad,apsw},
the electric period $a(u)$ does not become zero. Therefore, the $W$--boson 
and the quarks never become massless in the moduli space. I.e. the 
$W$ bosons are heavy everywhere in the moduli space and
the $SU(2)$ is always broken to $U(1)$ thus allowing for monopole solutions
in the whole moduli space.

Both $a_D$ and $a$ are solutions of a PF system. Therefore, finding
this curve immediately translates to the condition for the curve
being a solution of the underlying Schwarzian differential equation
with a proper choice of boundary conditions at $u=u_j$. 
For the massless case the curve is a solution of an usual 
Schwarzian differential equation \cite{matone,bf2}. 
This curve has been determined\footnote{See also \cite{henning}.} 
for the massless cases in \cite{afs,bf1,bf2}.
In the massive
case one has to solve a Schwarzian differential 
eqution of third order involving fifth order
derivatives which is quite involved \cite{schw}. In addition, a further problem
arises from the possiblity for the periods to have residua.
They appear in \req{bpsmass} in the last term.
Then the monodromie group is no longer $SL(2,\Z)$, but acts also non--trivially
on the quantum numbers $S_i$. E.g. for $N_f=1$, this has the effect that 
at $u\sim m^2 \ra\infty$, where the quark becomes massless, 
the weak--coupling monodromie transforms 
the monopole $(1,0)_S$ into $(1,0)_{S-1}$ .
Since $S$ is bounded, the monopole becomes a multiparticle state.
Therefore a jump in the spectrum should also take place at weak--coupling 
\cite{sw2}. On the other hand, since $u\sim m^2$ becomes 
a strong--coupling singularity as $m\ra 0$, which lies 
on the curve of marginal stability, it is obvious that `a part of that curve'
moves toward infinity.

Let us make a list of all states to be {\em expected} in N=2 supersymmetric
theories from consistency considerations, which basically follow from looking 
at the weak--coupling monodromy. For pure $SU(2)$ SYM one 
obtains\footnote{The electric charge $n_e$ is normalized
such that the charge of the quarks becomes integer. Then $a$ has 
to be divided by $2$ with the result that the mass \req{bpsmass} 
does not change. The quantum numbers refer to a certain choice of the 
basepoint, i.e. a specific choice of monodromie representation.} 
$(n\in\Z)$ \cite{sw1}

\be\label{bps0}
\begin{tabular}{|c|c|c|}\hline
         & {\rm weak--coupling} &  {\rm strong--coupling}\\ \hline\hline
         &                      &                        \\[-1.5mm]
$N_f=0$  &$ (0,2),\ (1,2n)$     &$ (1,0),\ (1,2)$        \\[2mm] \hline
\end{tabular}
\ee
\bc
{\em The expected spectrum in pure $SU(2)$ SYM.}
\ec
\ \\
Of course, all tables have to be completed with the corresponding 
anti--particles.
For $SU(2)$ SYM with $N_f=1,\ldots,3$ matter fields $(n\in\Z)$
the following weak--coupling spectrum is assumed\footnote{The quantum
numbers refer to the unbroken flavour group $SO(2)\simeq U(1),$\\ 
$Spin(4)\simeq SU(2)\times SU(2),\ Spin(6)\simeq SU(4)$ for $N_f=1,2,3$,
respectively \cite{bf2}. The index $6$ is \\
w.r.t. $SO(6)$.} \cite{sw2}:

\be\label{bpsweak}
\begin{tabular}{|c|c|c|}\hline
          & $m=0$                   & $m\lra \infty$          \\ \hline\hline
& &\\[-1.5mm]
$N_f=1$   & $(0,1)_1,\ (0,2)_0,\ (1,2n)_\h,\ (1,2n+1)_{-\h}
$ & $(0,1),\ (0,2),\ (1,2n)$\\[2mm] \hline
& &\\[-1.5mm]
$N_f=2$   & $(0, 1)_{(2,2)},\ (0,2)_0,\ (1,2n)_{(2,1)},\ (1,2n+1)_{(1,2)}
$ & $(0,1),\ (0,2),\ (1,2n)$\\[2mm] \hline
& &\\[-1.5mm]
$N_f=3$   & $(0,1)_6,\ (0,2)_0,\ (1,2n)_4,\ (1,2n+1)_{\bar 4},\ (2,2n+1)_0$ 
& $(0,1),\ (0,2),\ (1,2n)$
\\[2mm] \hline
\end{tabular}
\ee
\vspace{-0.7cm}
\bc
{\em The expected weak--coupling spectrum in $SU(2)$ SYM with $N_f$ flavours.}
\ec 
The table represents two limits $m=0$ and $m\ra\infty$.
The limit $m\ra\infty$ corresponds to the case where 
the quarks can be effectively integrated out from the theory with the result
of pure SYM \req{bps0}.
To determine the spectrum for $m=0$ one looks at the decomposition of the 
$Spin(2N_f)$ flavour quantum numbers when the mass goes from $m=\infty$ to 
$m=0$. These two limits, the classical limit $\La\ra 0$,
an $\Rc$--symmetry and the form of the instanton 
corrections are enough to determine the curves and therefore
the periods $a(u)$ and $a_D(u)$ in \cite{sw2}. 
By that one ends up with the requirement for existence of
special dyons with special $SO(2N_f)$ chirality.
The existence of these dyons with the specific $SO(2N_f)$ chirality
was proved in \cite{zaslow} for N=2 supersymmetric theories. 
In the $N_f=3$ case the dyon $(2,1)_0$ is required to exist and to 
become massless at a certain point $u_j$, which lies, along our previous 
arguments, on the curve of marginal stablity. 
Therefore this state should exist even at weak coupling. Then the 
weak--coupling mondromie can create the whole set of states $(2,2n+1)_0$.

As we will see later, the change in the spectrum from $m=\infty$ to $m=0$
goes continuously rather than discontinuous. 
More precisely, when
going from $m=0$ to $m=\infty$, the higher the mass becomes the more dyons
with odd electric charge will disappear from the spectrum.
Another quite non--trivial result is that at the singularity $u_j$
dyons with non--vanishing magnetic quantum numbers have to become massless
for small masses $m$ whereas for large $m$ the quarks should become massless.
We will comment on this issue later.
For the strong--coupling regime one expects \cite{sw2}

\be\label{bpsstrong}
\begin{tabular}{|c|c|c|c|}\hline
&  $m=0$&$m\neq 0$& $m \lra\infty$\\ \hline\hline
& & & \\[-1.5mm]
$N_f=1$   & $(1,0)_\h,\ (1,-1)_{-\h},\ (1,-2)_{\h}$& $(1,0),\ (1,-1),\ (1,-2)$
& $(1,0),\ (1,2)$\\[3mm] \hline
& & & \\[-1.5mm]
$N_f=2$   & $(1,0)_{(2,1)},\ (1,1)_{(1,2)}    $& $(1,0),\ (1,1)   $
 & $(1,0),\ (1,2)$\\[2mm] \hline
& & & \\[-1.5mm]
$N_f=3$   & $(1,0)_4,\ (2,-1)_0  $ & $(1,0),\ (2,-1)  $&
 $(1,0),\ (1,2)$\\[2mm] \hline
\end{tabular}
\ee
\bc
{\em \hspace{-1cm} 
The expected strong coupling spectrum in $SU(2)$ SYM with $N_f$ flavours}.
\ec
\ \\
Let us briefly mention the $N_f=4$ case. For vanishing bare
masses this is a scale--invariant theory without any quantum correction
to \req{bpsmass}. This means that the full $SU(2)$ is restored at $u=0$.
The condition \req{curve} is never satisfied implying that there is no 
curve of marginal stability, i.e. no distinction between strong-- and 
weak--coupling.
Since in this case the full $SL(2,\Z)$ acts 
on $(n_m,n_e)$ we can create the hypermultiplets $(p,q)$ and 
the vectormultiplets $(2p,2q)$, with 
$p,q$ relatively prime, from the elementary states $(0,1)$ and $(0,2)$,
respectively. Like in N=4 \cite{sen} these states appear as multi--monopole
bound states. 

The validity of a part of the above tables, namely for $m=0$,
was shown in \cite{bf1,bf2}. This was done by using the global discrete 
symmetry $\Z_{4-N_f}$, arising from a combination of an $\Rc$--symmetry and
a parity transformation on the $Q,\tilde Q$ and showing that in 
the strong--coupling regime, stable BPS states appear in 
$\Z_{4-N_f}$--multiplets \cite{bf2}. This discrete symmetry
acts on the Coulomb branch of the moduli space.
Besides the existence of a curve 
of marginal stability with the property \req{bound} was assumed.
Unfortunately, for $N_f=3$ such $\Z$--symmetry does not exist
and not much can be said about the $(2,-1)$--state.
Since in the massive case the $\Z_{4-N_f}$ symmetry is broken by the second 
term in \req{superW} and the geometry
of the curve of marginal stabilty is by no means obvious, such arguments 
can no longer be applied and one has to find new techniques.

In this work we want to address this question from a string point of view.
But string theory is not just a useful technique to answer
this question but it seems also to be the right 
framework for complicated scenarios which involve non--trivial interactions
between different types of dyons, as described before: the decay
of a photon into several other dyons at the curve of marginal stability 
or at the superconformal point the appearance of a monopole and a dyon 
becoming massless. We will see that the string picture puts these dyons
on the same footing as the gauge boson.

Towards this direction, the first step is to find 
the proper typeII string compactification, which 
has to be a $K3$--fibration with ADE--type singularity \cite{kklmv,klmvw}.
Such manifolds allow Riemann surfaces of the kind discussed in 
\cite{sw2} to emerge
naturally after taking proper limits. 
This fact implys that we can obtain all properties
of the Seiberg--Witten theories, like the periods and the BPS--spectrum
from that particular string vacuum. This was acomplished in the
case of pure $SU(2)$ for the periods
and monodromies in \cite{kklmv} and for the BPS--spectrum in \cite{klmvw}.
In this article we want to generalize this idea to the case of
$SU(2)$ SYM with matter fields. We want to verify the tables
\req{bpsweak} and \req{bpsstrong} from the string point of view.
Coming from six--dimensional Calabi--Yau compactifications
a BPS state from the tables above appears as a self--dual string 
on that Riemann surface $\Sigma$.
The question of stability of this state reduces to the question
of how the string winds around the SW--torus $\Sigma$, which
translates into the problem of looking for geodesics on this surface.

\sect{SW--geometry from ALE--fibrations}

One of the most striking manifestations of the recently discovered string
dualities is the equivalence of heterotic string compactified on $K3 \times
T^2$ with type IIA on a Calabi-Yau threefolds \cite{kv}.
The latter have to be $K3$ fibrations \cite{klm,al}. These are manifolds
with a $K3$ fibred over a base $CP^1$.
These theories have
N=2 supersymmetry in four dimensions and since the heterotic string theory
gives rise to perturbatively enhanced gauge symmetries we expect some
form of Seiberg--Witten theory to emerge in the dual type II picture. 
For large base space/weak coupling
the theory should behave like a $K3$ compactification of type IIA 
theory which on the other hand is dual to the heterotic string on 
$T^4$ \cite{ht}. 
The enhanced gauge
symmetries of the heterotic Narain lattice are reflected in ADE 
singularities on the $K3$. Furthermore, for finite base space 
non--perturbative effects should
deform this semi-classical gauge theory into the Seiberg--Witten theory.
This was demonstrated for some examples in \cite{kklmv}.
There it was shown that
in certain regions of the CY moduli space the CY periods turn into SW periods
dressed with gravitational $\alpha^\prime$--corrections. For this to work also
in more general cases one expects that the CY is a $K3$--fibration and the $K3$
fiber has an ADE--singularity somewhere on the moduli space. This means that
in a certain local neighbourhood the CY on the type IIB side
looks like a fibration of an ALE--space \cite{klmvw,lw}:

\be\label{ale}
W^\ast = \epsilon \left[ f(z,\La,..) + P^{ALE}_{ADE}(u_k,x,z_1,z_2) 
\right] + 
O(\epsilon^2) .
\ee
$P^{ALE}_{ADE}$ denotes the non compact ALE--space corresponding to the 
relevant
singularity of ADE--type, $z$ is the coordinate of the base $CP^1$ and
$\La\sim e^{-S}$ is the quantum scale. 
The precise form of the function $f$ will determine
how the $K3$ is fibred over $CP^1$. For the case of an 
$A_n$--singularity\footnote{The case for $E_6$ has been solved recently in 
\cite{lw}.}, which gives $SU(n+1)$ gauge symmetry, the relevant ALE space 
is given by \cite{klmvw}

\be\label{alespace}
P^{ALE}_{A_n} =2P_{A_n}(x;u_j)+z_1^2+z_2^2= x^{n+1} - u_2 x^{n-1} 
- \ldots - u_{n+1} + z_1^2 + z_2^2\ ,
\ee
where the $u_j$ denote the gauge invariant Casimirs and the quadratic pieces
are irrelevant deformations (from the singularity point of view). 
For $f$ we have the following expression:

\be\label{f}
f = z + \frac{\La^{2(n+1)}}{z}\ .
\ee
Up to the quadratic pieces in $z_1$ and $z_2$ the local, non--compact, piece
of the CYM \req{ale} 

\be\label{SWgeo}
z+\fc{\La^{2(n+1)}}{z}+P_{ADE}^{ALE}(u_k;x,z_1,z_2)
\ee
describes the curve for N=2 SYM with gauge group
$SU(n+1)$ without matter \cite{mw}.

The addition of matter fields results in a change of the base $f$.
For details see ref. \cite{klmw}. 
As in the case without matter \req{ale}, one expands the relevant 
CYM $W^\ast$ of type IIB around the $SU(2)$ symmetry enhancement point.
Then the lowest expansion in $\epsilon$ provides us
after some redefing
and rescaling of the CY moduli and coordinates a polynomial
whose $J$--function agrees with the one of the corresponding
SW--curves. 
In particular one obtains:

\be\label{finalSW}
W_{SW}=\lf\{ \ba{clccr} 
            & \ds{\fc{\La^4}{z} +z+ x^2-2u=0}&,&N_f=0\\[4mm]
            & \ds{\fc{\La_1^3}{z}+16 mz +8 z^2-x^2+8u=0}&,&N_f=1 \ea\ri.\ .
\ee

From integrating the holomorphic three--form $\Om$ of the CYM
over the relevant two--cycles of the $K3$, one can deduce
the periods $\la_{SW}$ of the Riemann surfaces \req{finalSW}.
One has to isolate the non--trivial two--cycles on the $K3$ 

\be\label{diff0} 
\int_{\nu_3}\Om \lra\int \fc{dx\ dz}{z}\ ,
\ee
to arrive at \cite{klmvw,klmw}

\be\label{diff}
\la_{SW}= x \frac{dz}{z}\ ,
\ee
with $x$ being the difference of any pair of branchpoints
of \req{finalSW}. 
The other two cases $N_f=2,3$ can be treated along these lines  \cite{klmw}.

\sect{BPS states as self--dual strings}

Let us describe how the BPS states of these supersymmetric field 
theories \req{finalSW} arise, which one obtains from expanding the CYM of 
type IIB around 
a $K3$--singularity and taking the limit $\ep\ra 0$.
The latter limit switches off gravity $\al'\ra 0$. 
The relevant BPS--states of typeIIB in ten dimensions correspond to
3--branes. Their space--time coordinates couple to a three--form 
with self--dual field--strength $C_{MNPQ}$. These 3--branes 
are wrapped around 3--cycles of the CYM.
There are two ways of wrapping these 3--branes over 3--cycles.
One possibility is that the 3--cycle has the topology of
the direct product $S^2\times S^1$
of a two--sphere $S^2$ lying in the ALE--space with a self--dual harmonic
two--form $g_{ij}$ near the singularity and a one--cycle $S^1$
lying on the curve \req{f}. In the other case the three--cycle $S^3$
cannot be decomposed in such a way. Nevertheless, in both cases
it should be possible to obtain a piece which can be related 
to a 2--cycle of $K3$ after taking proper projections.
The decomposition of the self--dual four--form
$C_{ij\al\beta}=g_{ij}\wedge B_{\al\beta}$ gives rise to an anti--symmetric 
tensor $B_{\al\beta}$ in six dimensions with self--dual field--strength. 
This $B$--field appears in a tensor multiplet with five other scalars.
The low--energy theory of type IIB on $K3$ is therefore described by a free 
tensor multiplet of N=2 and a light string
which couples to this $B$--field.
Since the field--strength of $B$ is self--dual 
this string is called self--dual.
As we will see later, its mass scales with the volume of the two--cycle
around which we wrap the three--brane.
This volume becomes small around the ADE--singularity.
Therefore the mass can be tuned to be much lighter than the string 
modes, i.e. it is a non--critical string without gravity.
It has N=2 supersymmetry in six dimensions 
and was first studied in \cite{witten}.
Sometimes it is also called tensionless string, since its tension also 
scales with the volume of this two--cycle.

Upon further compactification on \req{f} 
the $B$--field splits into $B_{\al\nu}=\om_{\al}A^{(4)}_\nu$.
The tensor--multiplet gives rise to a vector multiplet for the gauge field
$A^{(4)}_\nu$ in four dimensions. 
It may be electric or magnetic depending on the 
chosen homology cycle $\om_\al$ of the base.
Finally, wrapping the 3--brane, we have already wrapped around the 
vanishing 2--cycle of the $K3$, around the remaining $S^1$ we end up
with the self--dual string on the base $z$.

There is an important fact which arises due to the non--trivial
dependence on the base point $z$.
The local description \req{SWgeo} of the CYM $W^\ast$ can be written like:

\be
\prod_{i=1}^{n+1}[x-a_i(z)]+z_1^2+z_2^2=0\ .
\ee
In particular, from this form one immediately realizes \cite{bsv} that a 
vanishing two--cycle $S^2_{ij}$ is related to a point on the base $z$, with 
$a_i(z)=a_j(z)$, i.e. a point, corresponding
to the branchpoints $e_i$ of the curves \req{finalSW}, 
where the $K3$ degenerates.
In general, moving along a curve in the base $z$ implies 
a Weyl transformation on the set of two--cycles.
To avoid these transformations one introduces the Riemann surface $\Si$,
on which curves correspond to curves on the $z$--plane with trivial
action on the two--cycles. However this Riemann surfaces $\Si$ of genus $n$ 
are precisely the curves \req{finalSW} one obtains from the local 
fibrations up to the quadratic pieces in $z_1,z_2$: 

\be
\Si\ :\ \prod_{i=1}^{n+1}[x-a_i(z)]=0\ .
\ee
The other two branchpoints are $e_0=0,\ e_\infty=\infty$.
This tells us that the SW--torus appears on the typeII side as an object
on which one--cycles arising from wrapping three--branes around 
three--cycle are disentangled from the two--cycles of 
the local ALE--space, in contrast to the base $P^1$.

To make this more precise we take as example $N_f=1$ of \req{finalSW}.
The Riemann surface $\Si$ is

\be
\Si\ :\ [x-a_1(z)]\ [x-a_2(z)]=0\ ,
\ee
with:
\be\label{van2}
a_{1,2}(z)=\pm \sqrt{\fc{\La_1^3}{z}+16 mz+8 z^2+8u}\ .
\ee
Therefore we obtain a vanishing two--cycle for $a_1(z)=a_2(z)$, i.e. $z=e_i$
with the $e_i$ being the branchpoints of the base, i.e. solutions of

\be\label{zerosei}
\fc{\La_1^3}{z}+16 mz+8 z^2+8u=0\ \ \ ,\ \ \ z=e_1,e_2,e_3\ .
\ee

It was further explained in \cite{klmvw} that on this Riemann surface all
information about the three--cycles is encoded: E.g.: a three--cycle
$S^3$ on the CYM will correspond to a one--cycle on $\Si$, however to an open
curve that ends on two brachpoints $e_i$ on the base $z$. 
Moreover, after \cite{andy} three--branes wrapped around such three--cycles  
correspond to hypermultiplets.
On the other hand, a $S^2\times S^1$ corresponds to a closed curve on the 
base $z$ and also to a closed curve on $\Si$. A three-brane
wrapped around such three-cycles may give rise to vector-- or hypermultilpets
\cite{bsv}. In the cases under consideration they give rise 
to vector multiplets.

As we have explained before, the relict of a three--brane wrapped around a 
two--cycle $S_{ij}$ is a self--dual string, denoted by $ij$, on the $z$--plane. 
After \cite{bsv} to this two--cycle the pair $x=a_i(z),a_j(z)$ can be
associated. This means that \req{diff} is a multi--valued function
on $z$, however single--valued on $\Si$.
To determine the mass for this string, stretched between $z$ and $z+dz$, 
we have to use \req{diff} and take into account this ambiguity in $x$
with the result \cite{klmvw}:

\be\label{MASS}
M^{N_f}_{ij}(z)=\lf|[a_i(z)-a_j(z)]\  \fc{dz}{z}\ri|\ .
\ee
For our example \req{van2} we obtain from \req{diff}

\be\label{diff1}
\la_{SW}^{N_f=1}=2 \fc{dz}{z}\ \sqrt{\fc{\La_1^3}{z}+16 mz+8 z^2+8u}
\ee
and

\be\label{MASS1}
M^{N_f=1}_{12}(z)=2\lf|\sqrt{\fc{\La_1^3}{z}+16 mz+8 z^2+8u}\ \fc{dz}{z}\ri|\ .
\ee

We are interested in BPS states which finally represent the SW--spectrum.
Therefore we should wrap the three--branes around
minimal--volume three--cycles. This turns into the problem of 
minimising the mass of the string over the $z$--plane since the 2--cycle
over $z$ has already a minimal surface.
This translates into the condition of looking for geodesics on the $z$--plane.
Then, by construction, they locally minimize the mass \req{MASS}.
Following the arguments above, we should get a closed string (geodesics)
for a vector multiplet such as the $W$ boson. For hypermultiplets 
we expect open strings (geodesics) with their ends at the branchpoints $e_i$.
In particular the quarks, the monopole and all dyons should appear
in this way.
To conclude, what we are looking for are either geodesics running 
on the $P^1$--base between two
branchpoints $e_i$ which correspond to hypermultiplets or closed geodesics
running around a pair of branchpoints which represent
vectormultiplets i.e. gaugebosons. 
Of course, not all geodesics can be interpreted as stable BPS--states, i.e.
not all three--cycles will lead to stable BPS--states.
Since the meromorphic differentials $\la_{SW}$ have poles (e.g. $e_0=0$
in the $N_f=1$ case),  
there are geodesics going to a singularity.
Eq. \req{MASS} then tells us that this state
has infinite energy or mass. Therefore it cannot be a stable BPS state.
The same may happen for geodesics going to $e_\infty=\infty$.
It is quite amusing how the question of existence and stability
of BPS states in N=2 supersymmetric field--theories can be reduced
to the question of how three--branes of typeIIB are wrapped around 
three--cycles.

Let us briefly describe the situation for N=4 SYM one can extract
from type IIB on $K3\times T^2$ \cite{witten}. In contrast
to \req{MASS} the string mass
does not depend on the base due to the global product structure
$K3\times T^2$. It is given by

\be
M=\fc{\ep R}{\la_B}\ ,
\ee
with $\la_B$ being the string coupling of ten--dimensional type IIB.
The three--brane is wrapped around a two--cycle of $K3$ with volume
$\ep(z)=\ep\sim a_i-a_j$, 
now independent on the base, and a one--cycle with radius $R$
of $T^2$.
This formula has to be compared with \req{MASS}. In particular
reducing $\ep$
allows us to globally tune the mass
of all the BPS--states far below $M_{Planck}$ and other string excitations. 
By that one may derive the N=4 spectrum of supersymmetric gauge   
theories from the three-branes in ten dimensions, which become
self--dual strings on $T^2$ after wrapping the three--branes around 
the two--cycles of $K3$.
The BPS condition for the 3--cycles, namely  to have minimal volume,
translates to the condition for the geodesics to be just straight
lines on $T^2$. Therefore the W--bosons, monopoles and dyons are 
represented as strings going along the two cycles of $T^2$.
This gives us the general mass formula, valid for all combinations
of cycles, i.e. dyons $(p,q)$ on $T^2$ \cite{lll}

\be
M^{N=4}_{(p,q)}=\fc{\ep R}{\la_B} |p+\tau q|\ .
\ee
The complex structure modulus of $T^2$ becomes the coupling constant 
$\tau$. Then $SL(2,\Z)_\tau$ directly follows from $T$--duality of $T^2$
\cite{witten}. In \cite{lll} also the multiplicity\footnote{See also 
\cite{porrati2}.} of $(p,q)$--dyons following
from the $S_3$--action of $SL(2,\Z)$ on the $Spin(8)$ representations 
was checked. Purely field theoretical checks of the multiplicity of
dyons were done in \cite{porrati}.

In the following we want to discuss the BPS--spectra of the 
three cases $N_f=1,2,3$ in more detail.

\ \\
\underline{$N_f=1:$}
\ \\ 
\ \\
Instead of the curve \req{finalSW} we will now use the following form:

\be\label{newcurve}
W=\fc{\La_1^3}{z}+z^2 + 4 z m + 4 u\ ,
\ee
which has the same $J$--function. Eq. \req{diff1} now becomes:

\be\label{diff1a}
\la^{N_f=1}_{SW}=
\fc{dz}{z}\ \sqrt{\fc{\La_1^3}{z}+z^2+4zm+4u}\ .
\ee
As explained before, we have to search for self--dual strings 
running on the $z$--plane from the three branchpoints 
$e_i(u,m),\ i=1,2,3$, corresponding to solutions
of $\fc{\La_1^3}{z}+z^2 + 4 z m+ 4 u=0$.
The metric on $P^1$ is flat and the geodesics
are solutions of the following first order non--linear differential equation:

\be\label{nf1de}
\sqrt{\fc{\La_1^3}{z} +z^2 + 4 m z + 4 u}\  \frac{1}{z}\  
\ \frac{dz(t)}{dt} =\sqrt 2 Z\ ,
\ee
with (cf. eq. \req{bpsmass} )

\be
Z = n_m a_D + n_e a + \fc{1}{\sqrt 2}\sum S_i m_i\ .
\ee  
The term in the central charge
proportional to the mass $m$ arises from a residuum of the meromorphic
one--form $\lambda^{N_f=1}_{SW}$ located
at $z = \infty$. As we will see later, the latter are quite 
important to guarantee the
existence of geodesics for certain BPS--states. The initial condition is
$z(t = 0) = e_i $ for suitably chosen $i$.
It has been stressed already in \cite{klmvw}, that it is crucial to use
this special form for the differential \req{diff1}, i.e. \req{diff1a},
which follows from the CY \req{diff0}, 
since only that leads to the correct differential equation for the geodesics.
Modifications, which still give the same periods, 
will change the differential equation.

The discriminant $\Delta_1(u,m)$ is a polynomial of degree 3
in $u$. Therefore we expect three dyons $(1,0),\ (1,1)$ and $(1,2)$,
to become massless at the zeros $u_1(m), u_2(m)$ and $u_3(m)$, respectively.
This implies a collapse of branchpoints
\bea
e_1&\lra& e_3\ \ \ ,\ \ \  u\lra u_1(m)\nnn
e_2&\lra& e_3\ \ \ ,\ \ \  u\lra u_2(m)\nnn
e_1&\lra& e_2\ \ \ ,\ \ \  u\lra u_3(m)\ , 
\eea
respectively.
For $m=0$ these three singularities are $u_1=-3\cdot 2^{-\fc{8}{3}},\ 
u_2=3\cdot 2^{-\fc{8}{3}} e^{-\fc{\pi i}{3}}\ ,
u_3=3\cdot 2^{-\fc{8}{3}} e^{\fc{\pi i}{3}}$.

\begin{itemize}

\newpage
\item \underline{$u = -1,\ \ m = 0$}:
\vspace{-1.5cm}
\begin{figure}[h]
\hbox to\hsize{\hss
\epsfysize=16cm
\epsffile{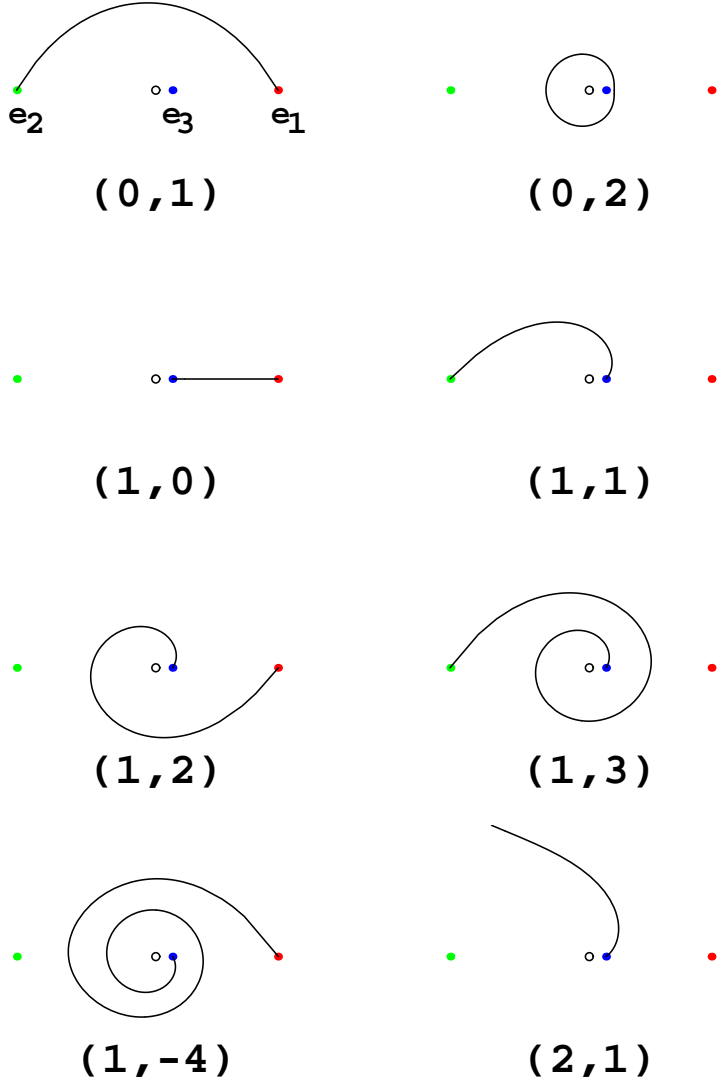}\hss}
\vspace{-0cm}
\caption{$N_f = 1$ , $u = -1$ , $m = 0$}
\label{bps12}
\end{figure}

In this picture we have depicted the results  of our analysis.
Stable states correspond to closed
trajectories or open trajectories running between two
branchpoints. Trajectories that run off to infinity  or
to zero correspond to unstable states.
The string representing e.g. the monopole goes from the branchpoint
$e_1$ to $e_3$. These are the branchpoints which collapse at
the singularity $u=u_1$. On the CYM this then corresponds
to a vanishing three--cycle around which the three--brane is wrapped.
All dyons $\pm(1,n)$ and $\pm(1,2n')$ are present in agreement with table 
\req{bpsweak}.
The other two states, becoming massless at $u_2$ and $u_3$ show up 
in a very similar way. The gauge boson is represented by a closed 
(counterclockwise) geodesic around the branchpoints $e_3$ and 
the point $z=0$ which is
drawn as  a small circle in our pictures.
Thus we find the expected
weak coupling spectrum.
The trajectory for the $(2,1)$ state runs to infinity i.e. this
state is unstable.

\item\underline{$u = -1,\ \ m = 0.5$}:

\vspace{-3.2cm}
\begin{figure}[h]
\hbox to\hsize{\hss
\epsfysize=16cm
\epsffile{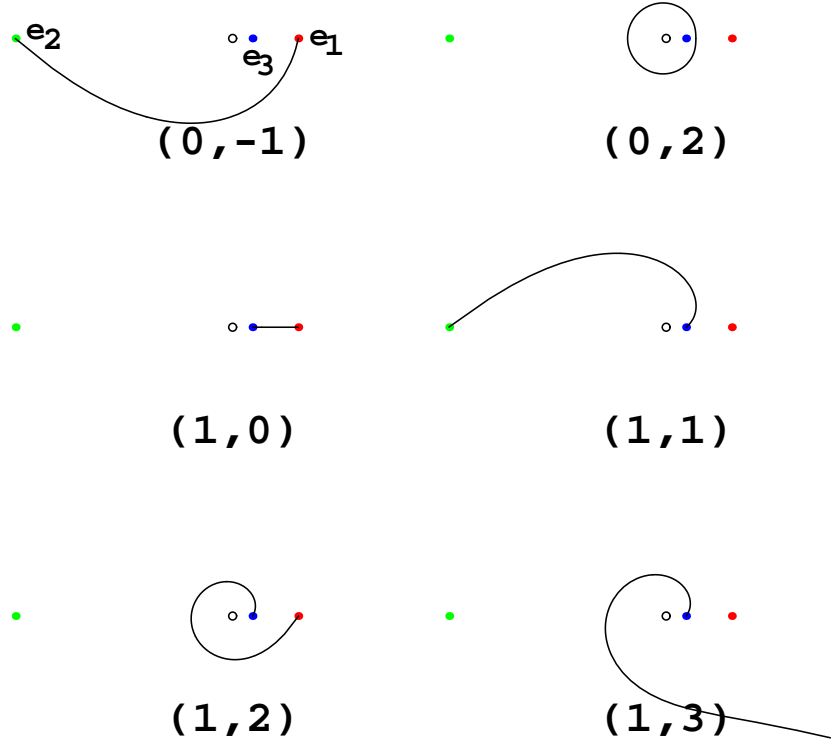}\hss}
\vspace{-1.2cm}
\caption{$N_f = 1$ , $u = -1$ , $m = 0.5$}
\label{bps13}
\end{figure}

In the massive case we encounter a residue of $\fc{2m}{2\pi i}$ at  
$z=\infty$. A residuum has the effect that the 
periods $a$ and $a_D$ are no longer
invariant when we deform the homology cycles across $z_R$.
Therefore, it modiefies the BPS mass formula and the equation
for the geodesics.  On the other hand, this fact is important  to guarantee
the existence of the strings $(1,0),  (1,1)$ and $(1,2)$ which are 
responsible for the singularities at $u_1(m),u_2(m)$ and $u_3(m)$, 
respectively, 
for non--vanishing masses. This is also precisley what we see in the picture 
and
without this modification, we would not obtain e.g.: the dyon $(1,1)$.
We also see that already some dyons with odd electric charge $(1,2n+1)$
disappear from the spectrum, i.e. in the string picture their geodesics
run to zero or infinite. We studied the spectra for various real values
of $m$ and  found that for increasing $m$ more and more $(1,2n+1)$
dyons disappear.  Only the $(1,1)$ dyon is present also for large mass. 
E.g.: the $(1,3)$ dyon is still present for $m=0.1$, but no longer in fig. 2
at $m=0.5$

Recently, in \cite{f} some  arguments were  
given, that $S$ is also a function of $u$ and differs from its value
in the massless case. The quark $S=1$ has period $a + m$ but
in our picture the state $(1,0)$ has period $a_D$ and not
$a_D+m/2$
which one would expect from the mass formula for the monopole
with $S=1/2$.
Similarly the periods for the $(1,2n)$ dyons is $a_D + 2 n a$.
On the other hand for
the $(1,1)$ state with $S=-1/2$ the period is $a_D+a-m$ and not
$a_D+a-m/2$. For the other states with odd electric charge the
period is $a_D+(2n+1)a-m$.
Thus it seems that part of the contribution of the residue
is already included
in $a_D$. This means that as we send $m$ to
infinity the monopole and the dyons with even electric charge survive
since $a_D$ and $a$ remain finite in this limit but the
$(1,1)$ dyon disappears
because its mass becomes infinite. The difference of the $S$ quantum numbers
is the same in the massive and the massless case but the absolute value
changes in the massive case. This supports the conclusion
that the quark number $S$ has to be distinguished from its physical
value that appears in the central charge formula \cite{f}.

\item\underline{$m \ra \infty$\ ,\ $\La_1\ra0$}:

As we increase the mass, we should finally end up with SYM with 
the matterfield being effectively integrated out.
From the curve \req{newcurve} written in the form:

\be
W=\La_0^2\lf[\fc{1}{\tilde z}+4 \tilde u+4 \tilde z+\tilde z^2 
\fc{\La_0^2}{m^2}\ri]-x^2\ ,
\ee
with $\tilde z=z \sqrt{\fc{m}{\La_1^3}}$\ , $\tilde u=\fc{u}{\La_0^2}$
and  $m\La_1^3=\La_0^4$ we obtain from \req{diff1a}

\be
\lambda_{SW}^{N_f=1}=2\La_0 \fc{d\tilde z}{\tilde z} 
\sqrt{\fc{1}{\tilde z}+4 \tilde u+4 \tilde z+\tilde z^2 
\fc{\La_0^2}{m^2}}\ .
\ee
From these expressions one recovers how pure SYM arises for $m\ra\infty$.
For the limit $m\ra\infty$ we obtain the $W$ and $\lambda$ of \cite{klmvw}
up to irrelevant rescalings which do not change the $J$--function.
Whereas $\tilde e_1=e_1 \sqrt{\fc{m}{\La_0^2}}$ and $\tilde e_3$ become 
the branchpoints of the SW--curve, expressed in $\tilde u$, namely

\be
\tilde e_{1,3}\lra -\fc{\tilde u}{2}\pm \h\sqrt{\tilde u^2-\La_0^4}\ ,
\ee
the third branchpoint $\tilde e_2$ moves to infinity:

\be
e_2\lra -\infty\ .
\ee
During that process we loose many stable BPS states, namely
all $(1,2n+1)$--dyons, which end at $e_2$ and become heavy.
The three singularities $u_1(m),u_2(m)$ and $u_3(m)$ become

\bea
\tilde u_{1,3}&\lra& \pm 1\nnn
u_2&\lra& m^2\nnn
\eea
in agreement with what one expects \cite{sw2}.
In particular this means that the quark becomes massless in the 
weak--coupling region at $u\ra u_2(m)=m^2$.

Looking at fig.1, at this limit $e_2 \lra e_3$ and one would expect the dyon 
(1,1) to become massless at this point.  
However, as noted in \cite{sw2} 
there is an interesting phenomenon happening when
we increase the mass from zero to higher values. What appears
as a dyon for small mass appears as an elementary particle for
large mass. This is due the non-Abelian monodromies since what we call
a monopole depends on the choice of a basepoint and a path around
the singularity in the moduli space.
When we start to increase the mass the singularities will move around
each other in moduli space
and the chosen path is deformed and will differ from the natural choice
of path we would make for large mass. When we change the choice of paths
the monodromy will be conjugated
while it still belongs to same conjugacy class. This conjugation changes
the magnetic and electric quantum numbers e.g. a dyon may become a quark.

\item \underline{$u = 0,\ \ m = 0$}:

\vspace{-1.7cm}
\begin{figure}[h]
\hbox to\hsize{\hss
\epsfysize=12cm
\epsffile{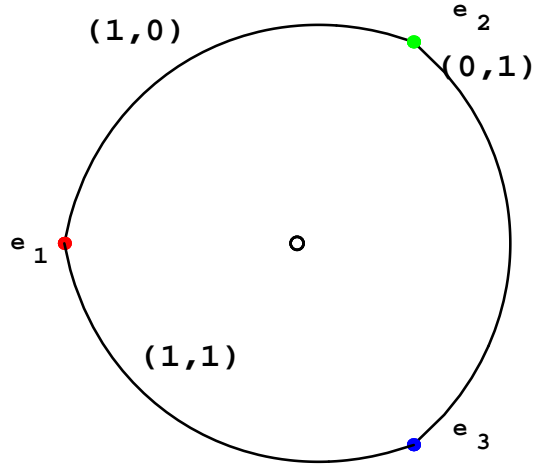}\hss}
\vspace{-3cm}
\caption{$N_f = 1$ , $u = 0$ , $m = 0$}
\label{bps11}
\end{figure}
Notify that the dyon $(1,2)$ shows up with the quantum numbers $(0,1)$ 
in the strong coupling region. This is because, in general, 
one has to relabel quantum numbers when crossing a branch cut 
of the periods to guarantee the BPS mass formlua \req{bpsmass}
to stay a smooth function in $u$ \cite{bf2}.
In other words, these states are related to the
conventional notation of the
strong--coupling states $\{(1,0),(1,1),(1,2)\}$,
which only refer to a certain choice of paths around the singularities 
and a choice of the basepoint. For the basepoint $u=0$ the
path for the $(1,2)$ state is not the natural choice.
The natural path around $u_3$ yields a dyon with charges
$(0,1)$ and the 
corresponding monodromy is related to the monodromy of
the $(1,2)$ state by conjugation.
It is the quantum number $S$ which says that this  state 
with electric charge one is rather a dyon $(S=\h$) than  a quark
($S=1$).
In this picture we also see very nicely the appearance of  a monopole and 
two dyons with 
mutually non--local charges `at the same time', i.e their monodromies do 
not commute.
As we have explained before,  on the CYM these states correspond to 
three--cycles with non--trivial intersection number.
At the curve of marginal stability the $W$--boson decays.
This decay, involving an interaction between particles with 
mutually non--local charges, can be recognized in the picture:
The sum of the paths of all three strings with magnetic charge
adds up to a closed circle, which includes  $e_0=0$.
Therefore, after  fig. \req{bps12} it must be identified with the $W$--boson.

\be\label{decay}
(0,2)= (1,1)+ (0,1) -(1,0)\ .
\ee
There is a minus--sign for the monopole, since this string goes
into the opposite direction, in contrast to all other states which
go counterclockwise, in agreement with the $W$--boson in  fig. \req{bps12}.
Thus, we see that the $W$--boson appears  `virtually' in the figure.
After multiplying eq. \req{decay} by two we recover the $W$--boson
decay which preserves the $S$--charge \cite{bf2}

\be\label{decay1}
(0,2)_0\lra 4(0,1)_{\h}+2(1,-1)_{-\h}+2(-1,0)_{-\h}\ .
\ee

\item \underline{$u = 0.3,\ \ m = 0.5$}:
\vspace{-3.5cm}
\begin{figure}[h]
\hbox to\hsize{\hss
\epsfysize=14cm
\epsffile{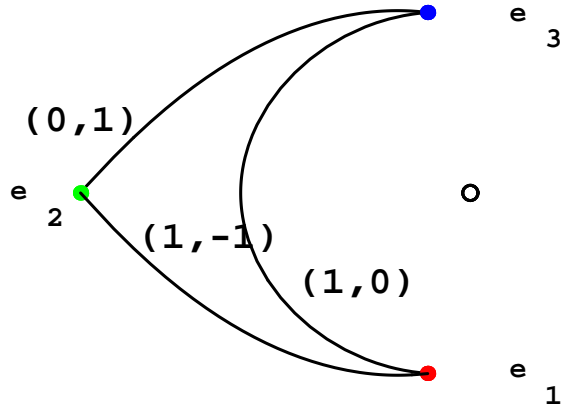}\hss}
\vspace{-3cm}
\caption{$N_f = 1$ , $u = 0.3$ , $m = 0.5$}
\label{bps14}
\end{figure}

In the massive case, we keep all three strong--coupling states, since we still
have the three singularities in the $u$--plane, which have moved
depending on the mass $m$.
This is in agreement
with table \req{bpsstrong}. We are now in a strong--coupling region
where it is not possible to obtain formally the $W$--boson loop
surrounding $e_0$. We have:

\be
(0,1)+ (1,-1) -(1,0)=(0,0)\ .
\ee
Notify, that we have really zero on the right hand side, since we also do not
encounter the residuum.

\item \underline{$u = 0,\ \ m = 0.75$}: 
\ \\
In the $N_f=1$ case we have a superconformal point at e.g.: $u=\fc{3}{4}=m$,
where two particles with mutually non--local charges become massless.
In this case these are the dyons $(1,1)$ and $(1,2)$ \cite{bf1,apsw}.
At such points all three branchpoints coincide, as it can already be
seen from fig. \req{bps19}. Since these two dyon--strings
run between two such branchpoints, from the string--picture it is
obvious that these particles become massless.
Since one is able to  approach this limit while keeping both states this could
be a hint that string--theory may be the right framework to describe
the physics near and at the superconformal points.
The periods vanish at this point. After \req{bpsmass} this indicates 
that the mass of the other BPS states is entirely given by the residuum. 
As in fig. 3 we can visualize the $W$--boson:

\be
(0,2)=(-1,0)+(1,1)-(0,-1)\ ,
\ee
allowing for the decay \req{decay1} with non--trivial interactions.
Again, the sign of $(0,-1)$ is reversed, since it
runs clockwise, in contrast to the other two dyons.
In this case we also pick up a residuum, when encircling along the $W$--boson
loop.

\vspace{-4cm}
\begin{figure}[h]
\hbox to\hsize{\hss
\epsfysize=14cm
\epsffile{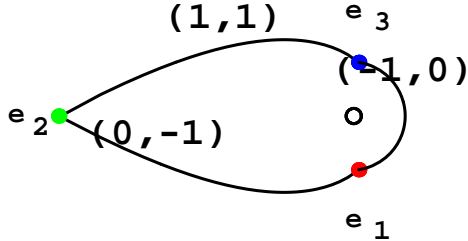}\hss}
\vspace{-4cm}
\caption{$N_f = 1$ , $u = 0$ , $m = 0.75$}
\label{bps19}
\end{figure}

\end{itemize}

\pagebreak
\ \\
\underline{$N_f=2:$}
\ \\
\ \\
For the $N_f=1$ case we have studied trajectories 
in the cut $P^1$--base. They are to be related to curves on the 
Riemann surface $\Si$. Therefore we can also investigate
the properties of the BPS--states on this surface directly.

\ \\
Let us now discuss the case with two flavours by
writing down the necessary ingredients to find the BPS--states.
The details about the relevant CYM can be found in \cite{klmw}.
The curve one ends up is:

\be
y^2=\lf(x^2-\fc{\La^4_2}{64}\ri)(x-u)+\fc{m^2\La_2^2x}{4}-\fc{m^2\La_2^4}{32}\ ,
\ee
where we have set $m_1 =  m_2 = m$.
In the $x$--plane we have the following three branchpoints:

\bea
e_1&=&\ds{\fc{\La_2^2}{8}}\nnn
e_2&=&\ds{\fc{u}{2}-\fc{\La_2^2}{16}-\fc{1}{16}
\sqrt{(\La_2^2-8u)^2-32\La_2^2(2m^2-u)}}\nnn
\nnn
e_3&=&\ds{\fc{u}{2}-\fc{\La_2^2}{16}+\fc{1}{16}
\sqrt{(\La_2^2-8u)^2-32\La_2^2(2m^2-u)}\ .}
\eea
The positions of the three strong coupling singularities are:

\bea
u_{1,2}&=&\ds{-\fc{\La_2^2}{8} \mp \La_2 m}\nnn
u_3&=&\ds{\fc{\La_2^2}{8} + m^2}
\eea
The meromorphic one-form is given by:

\be
\la^{N_f=2}_{SW}=-\fc{\sqrt{2}}{4\pi}\fc{dx y}{x^2 - \fc{\La_2^4}{64}}\ .
\ee
Notify that this one--form has a residue at $x_R=-\fc{1}{8}\La_2^2$.
For generic masses there is also a residue at  $x_R=\fc{1}{8}\La_2^2$
which vanishes when the masses are chosen to be equal.
The residue vanishes for the limit $m\ra\infty, \La_2\ra 0$ with
$m^2 \La_2^2 = \La_0^4$, in
agreement
with the fact that we do not expect any residuum in the pure SYM 
case \cite{sw2}.
Let us first discuss the weak--coupling region:

\newpage
\begin{itemize}

\item   \underline{$u = 1$,\ \ $m = 0$}:

\vspace{-2.5cm}
\begin{figure}[h]
\hbox to\hsize{\hss
\epsfysize=16cm
\epsffile{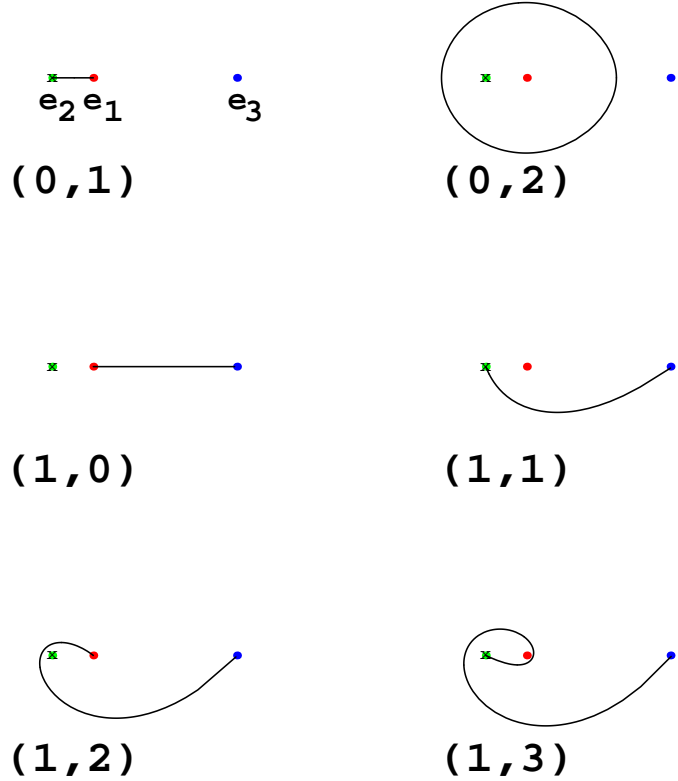}\hss}
\vspace{-1.5cm}
\caption{$N_f = 2$ , $u = 1$ , $m = 0$}
\label{bps21}
\end{figure}

As in the $N_f = 1$ case we  find the expected weak-coupling
spectrum \req{bpsweak}; the quark and the dyons are represented by open
trajectories
and the W-boson by a closed one. The cross in the $N_f = 2$
pictures denotes the branchpoint at $x = - \fc{1}{8} \La_2^2$ which,
in the  case of  non--zero mass $m$, coincides with
the  location of the residue. We were not able to find closed
geodesics  with magnetic quantum numbers greater than  one and
conclude that these states are  unstable as expected.

\newpage
\item   \underline{$u = 1$,\ \ $m = 0.7$}:

\vspace{-2.5cm}
\begin{figure}[h]
\hbox to\hsize{\hss
\epsfysize=16cm
\epsffile{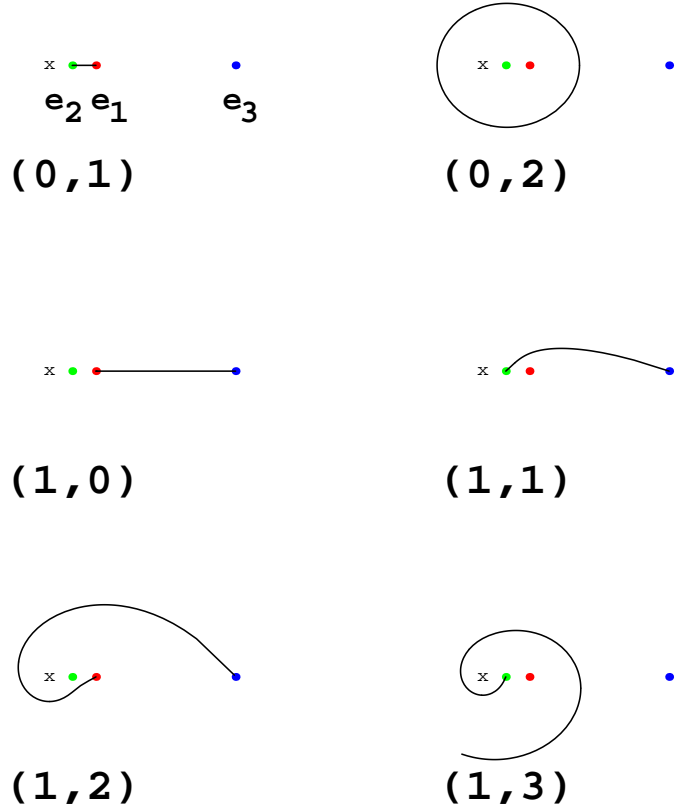}\hss}
\vspace{0cm}
\caption{$N_f = 2$ , $u = 1$ , $m = 0.7$}
\label{bps23}
\end{figure}

In the massive case the form of the periods for the BPS states
is the same as in the $N_f = 1$ case i.e.
we have $a_D + 2 n a$ for $(1,2n)$ dyons and
$a_D + (2n+1) a + m$ for the $(1,2n+1)$ dyons.
If we stay at weak coupling and turn on the mass, the higher the mass
becomes, the more $(1,2n+1)$ dyons with large elecric charge disappear.
In particular we see that the spectrum does not jump suddenly.
Only the $(1,1)$ dyon fig. \req{bps23} survives
also at large $m$ since it is responsible  for the singularity
at $u  = u_3$. In the flow to the pure gauge theory this singularity
goes  to infinity  and  the singularities at $u =  u_{1,2}$
become the monopole and dyon point of the pure SYM theory.

\item   \underline{$u = 0$,\ \ $m = 0$}:

As we decrease $u$ for $m = 0$ we will cross the curve of
marginal stability where all states decay into two states,
the monopole $(1,0)$ and the $(1,1)$ dyon. 

\newpage
\vspace{-10cm}
\begin{figure}[h]
\hbox to\hsize{\hss
\epsfysize=13cm
\epsffile{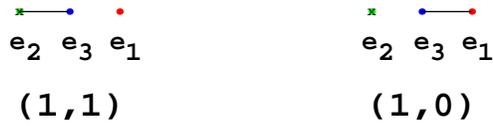}\hss}
\vspace{-5cm}
\caption{$N_f = 2$ , $u = 0$ , $m = 0$}
\label{bps24}
\end{figure}
In our analysis the strong coupling spectrum 
appears as two straight open strings
stretched between two branchpoints (with opposite direction).
See fig. \req{bps24}. Again, both states add up to the $W$--boson 
cycle 

\be
(0,2)\lra 2(1,1)-2(1,0)\ ,
\ee
after taking into account the different directions of the two strings.

\item   \underline{$u = 0$,\ \ $m = 0.7$}:

\vspace{-4cm}
\begin{figure}[h]
\hbox to\hsize{\hss
\epsfysize=13cm
\epsffile{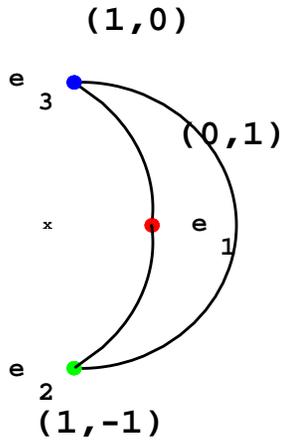}\hss}
\vspace{-3cm}
\caption{$N_f = 2$ , $u = 0$ , $m = 0.7$}
\label{bps25}
\end{figure}
On the other hand, when we turn on the mass, the monopole
singularity $u_{12}=-\fc{\La_2^2}{8}$ splits into two singularities
$u_{1,2}$ [cf. eq. (3.26)]
and therefore we expect  
three dyonic states in the strong coupling regime
 [e.g.: $(1,0),\ (1,1),\ (1,2)$].
After fig. \req{bps25} these three states add up to
zero, since we do not encounter the zero $e_0$ like the $W$--boson
in fig. (7):

\be
(0,1)+(1,-1)-(1,0)=(0,0)\ .
\ee
We also do not pick up a residuum.
This picture is similar to fig. (4).
\end{itemize}

\ \\
\underline{$N_f=3$}:
\ \\
\ \\
The curve for the theory with three flavors of equal mass
is given by \cite{sw2}:

\be
W  = x^2(x-u)-\fc{1}{64}\La_3^2(x-u)^2-\fc{3}{64}m^2\La_3^2(x-u)+
\fc{1}{4}m^3\La_3x-\fc{3}{64}m^4\La_3^2\ .
\ee
The differential takes the following form:

\be\label{diff3}
\la_{SW}^{N_f=3}=\fc{\sqrt{2}}{8\pi}\fc{dx\ x(3m^2+6m^3-4mu+2 mx-2
ux+x^2)}{(m^2-x^2)
\sqrt{-3m^4+3m^2-u^2-3m^2x+2m^3x+2ux-x^2-ux^2+x^3}}.
\ee
where we have set $\La_3=8$.
The three branchpoints are determined to ($m=0$)::

\bea
e_1&=&u\nnn
e_2&=&\ds{\h(1+\sqrt{1-4u})}\nnn
e_3&=&\ds{\h(1-\sqrt{1-4u})\ .}
\eea
They collapse $e_1\ra e_3$ for $u\ra 0$ with the monopole becoming massless
and $e_3\ra e_2$ for $u\ra \fc{1}{4}$ with the $(2,-1)$ 
dyon becoming massless, respectively.
Let us focus on two cases with zero mass, both at weak  coupling.
They differ by the arrangement of the branchpoints in the $x$--plane.
From these two examples, one can see how different the BPS--states  
are represented depending on the modulus. E.g.: in fig. 10
one would guess the quark becoming massless for $e_2\ra e_3=\h$.
However, in fig. 11, which is also the right patch to approach this limit,
one realizes that when $e_2,e_3\ra\h$ and $e_1\ra \fc{1}{4}$ the quark
encloses the branchpoint $e_1$, i.e. it cannot become massles.

\newpage
\begin{itemize}
\item   \underline{$u = -6$,\ \ $m = 0$}:
\begin{figure}[h]
\hbox to\hsize{\hss
\epsfysize=16cm
\epsffile{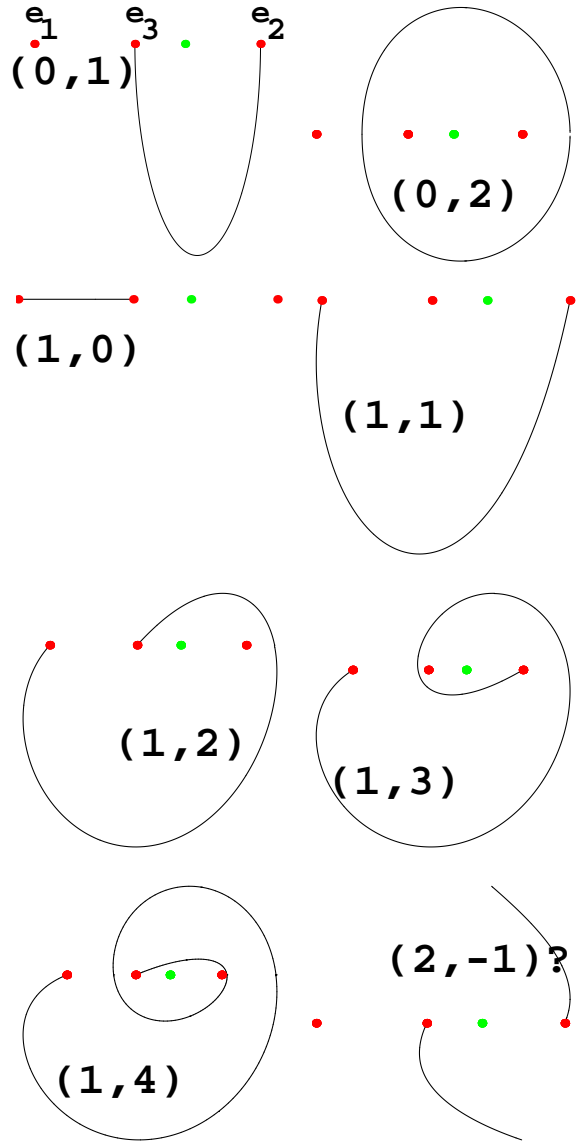}\hss}
\caption{$N_f = 3$ , $u = -6$ , $m = 0$}
\label{bps30}
\end{figure}

In this case we have: $e_1=-6,\ e_2=3,\ e_3=-2$.
The geodesics we found
are the quark, the  W-boson  and the $(1,n)$ dyons in
agreement with \req{bpsweak}. But the strings 
corresponding to states with magnetic
charge two predicted by \cite{sw2} could not be
found despite strong efforts fig.\req{bps30}.

\newpage
\item   \underline{$u = 1/2$,\ \ $m = 0$}:
\vspace{-1cm}
\begin{figure}[h]
\hbox to\hsize{\hss
\epsfysize=16cm
\epsffile{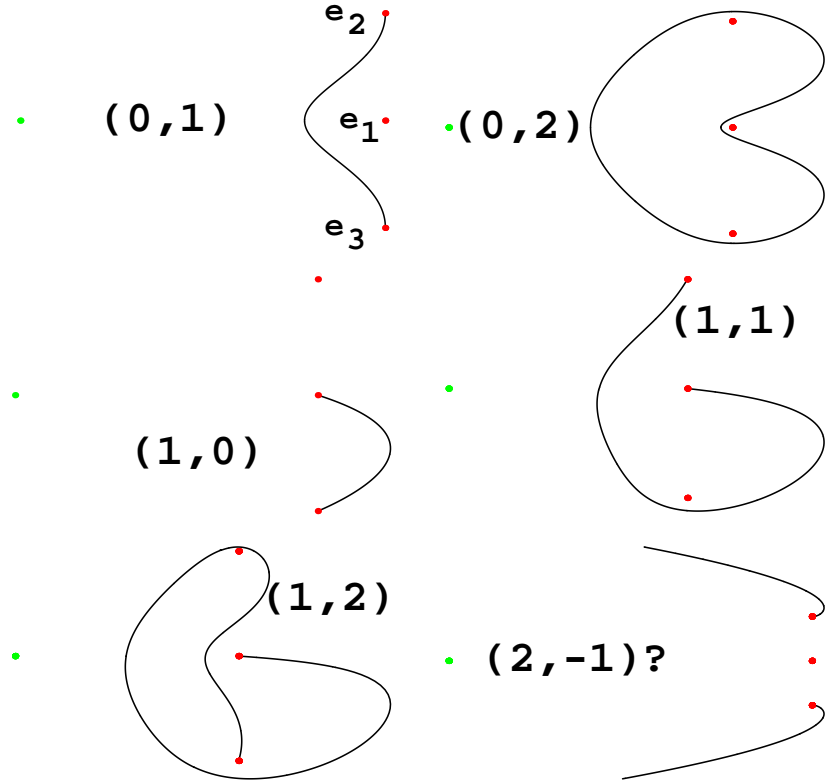}\hss}
\vspace{-2cm}
\caption{$N_f = 3$ , $u = \fc{1}{2}$ , $m = 0$}
\label{bps31}
\end{figure}
\end{itemize}
The situation is  similar to the $u=-6$  case,
the states with magnetic charge up to one were found
but in our pictures the $(2,-1)$ state has a non--expected behaviour:
For the case of $u=1/2$, it should be represented by a geodesics starting
at $e_2$, passing by $e_1$ and finally ending at $e_3$. This 
fact could be related that the meromorphic differential \req{diff3}
is not the canonical one.  

For the strong--coupling spectrum we obtain 
a similar figure as fig. 8 with $(1,1)$ just replaced by $(2,-1)$.

\newpage
\sect{Concluding remarks}

We have studied the spectrum and the stability 
of BPS states in N=2 SYM theory with $N_f=1,2,3$ massive flavours in the 
fundamental representation.
This has been accomplished by investigating a point--particle
limit of the underlying typeII string--compactifi--cation which
leads to the corresponding $SU(2)$ SYM theory with massive matter. The
BPS--states arise on typeIIB from three--branes in ten dimensions.
After wrapping them around the vanishing two--cycles of the local $K3$
they become self--dual strings on an Riemann manifold $\Si$ which is 
equivalent
to the SW--torus. We could establish a correspondence between self--dual 
strings on such manifolds and the field--theoretical BPS--states one expects 
from N=2 SYM. This enabled us to study their stability and strong--coupling
behaviour. Indeed we could verify the tables,
shown in the introduction, up to one discrepancy concerning
the $(2,1)$ state in the $N_f=3$ case.
In this picture gauge fields and monopoles appear
on equal footing, since the two homology cycles of $\Si$ correspond
to magnetically and electrically charged self--dual strings, respectively.
This picture seems to be the right framework to describe, e.g. 
particles with mutually non--local charges becoming massless or decaying
at certain regions in the moduli space.  
This is a quite difficult task in field theory where gauge bosons
appear as elementary particles and monopoles as solitonic solutions
or vice versa. 

Unfortunately, for the case of three flavours we could not find evidence for 
the existence of the $(2,1)$ state, which would be very important for 
consistency which follows from matching the $m\ra \infty$ spectrum with 
flavour symmetry $SU(3)\times U(1)$ (for equal quark masses) with the $m=0$ 
spectrum with $SU(4)$.
We interpret \cite{bf2} such that the existence of the dyon (2,1) was 
established on the ground of \cite{zaslow}, mainly because of the lack
of a $\Z$--symmetry in the moduli space. 
In global N=2 SUSY one can construct such a state \cite{zaslow}.
But it still remains to proof that the curve for the $N_f=3$ case also 
allows for such a state as it has been proposed in \cite{sw2}.

The SW--geometries and their BPS--spectra 
arise in \cite{kklmv,klmvw} and for the cases, 
we considered here, from higher dimensional theories with extended objects, 
namely type IIB with solitonic three--branes.
There are now many recent results where the curve
or the effective action of N=2 SYM without or with massive matter 
emerges either as useful tool
to describe some deformations away from an orbifold point
of $F$--theory on $K3$  \cite{sen05}, as 
effective theories which describe some higher dimensional
theories in a certain region of their moduli space \cite{ganor} 
or as part of a more complicated theory \cite{dl,bds} and e.g.
\req{bpsmass} arises as the mass of $(p,q)$--strings 
going between different branes \cite{sen08}.
Therefore it should be possible to obtain the results we have gotten here 
within those frameworks. I.e. starting from the BPS--states in these 
theories one should be able to deduce constraints
for the field--theortical Seiberg--Witten spectrum. 
This was done partially in \cite{sen08} based on the equivalence between
N=2 $SU(2)$ SYM with $N_f=4$ and the world--volume theory of a three--brane 
of type IIB in the presence of a configuration of four 7 D--branes and an 
orientifold plane \cite{bds}.
In this case, the mass formula of the BPS--states \req{bpsmass} can be 
related to the mass of an open string going from a three--brane to a 
seven--brane.
A BPS state is obtained when the mass is minimized, i.e. when the path
from the three--brane to the seven--brane goes along a geodesics.
Notify, that ref. \cite{ganor} obtains also the curves for $N_f=3$.
Geodesics have also appeared recently in a different context,
when discussing symmetry enhancement in $F$--theory on $K3$, where open
strings representing the $W$--bosons 
connect seven--branes along geodesics \cite{joh}.

\ \\ \\
{\em {\mbox{\boldmath $Acknowledgement:\ $}}}
We would like to thank W. Lerche and P. Mayr for
drawing our attention to the problems presented here. 
Moreover, we are very grateful
to
J.--P. Derendinger,
P. Mayr,
A. Klemm,
W. Lerche and S. Theisen 
for helpful discussions.
We also thank J.--P. Derendinger for providing excellent 
working conditions in Neuch\^atel and P. B\'eran for some Fortran 
assistance. The research of A. B. has been supported in part by the
Bundesministerium f\"ur
Wissenschaft, Verkehr und Kunst.



\begin{thebibliography}{99}

\newcommand{\np}{\mbox{\em {Nucl. Phys.} {\bf B }}}
\newcommand{\pl}{\mbox{\em {Phys. Lett.} {\bf B }}}
\newcommand{\cmp}{\mbox{\em {Comm. Math. Phys. }}}
\newcommand{\prd}{\mbox{\em {Phys. Rev.} {\bf D }}}
\newcommand{\prep}{\mbox{\em {Phys. Rep.\ }}}
\newcommand{\mpl}{\mbox{\em Mod. Phys. Lett.\ } {\bf A}}
\newcommand{\bi}[1]{\bibitem{#1}}

\bi{sw1} N. Seiberg and  E. Witten, \np {\bf 426} (1994) 19

\bi{sw2} N. Seiberg and  E. Witten, \np {\bf 431} (1994) 484

\bi{wl} For a review see e.g.: W. Lerche, {\em Notes on N=2 supersymmetric 
gauge theory}, Proceedings Gauge Theories, Applied Supersymmetry and 
Quantum Gravity, pp. 53--79, Leuven University Press (1996)

\bi{sen} A. Sen, \pl {\bf 329} (1994) 217


\bi{zaslow} S. Sethi and M. Stern, hep--th/9507145;\\
S. Sethi, M. Stern and E. Zaslow, hep--th/9508117;\\
J.P. Gauntlett and J.A. Harvey, hep--th/9508156

\bi{klt}  A. Klemm, W. Lerche and S. Theisen, {\em Int. J. Mod. Phys.}
{\bf A 11} (1996) 1929


\bi{bs} A. Brandhuber and S. Stieberger, hep--th/9609130;\\
Y. Ohta, hep--th/9604051 and hep--th/9604059;\\
T. Masuda and H. Suzuki, hep--th/9609066, hep--th/9609065

\bi{ito} K. Ito and  S. Yang, \pl {\bf 366} (1996) 165

\bi{bf2} A. Bilal and F. Ferrari, hep--th/9605101


\bi{ad} P.C. Argyres and M. Douglas, \np {\bf 448} (1995) 93

\bi{apsw} P.C. Argyres, M. Plesser, N. Seiberg and E. Witten,
\np {\bf 461} (1996) 71


\bi{matone} M. Matone \prd {\bf 53} (1996) 7354

\bi{henning} M. Henningson, \np {\bf 461} (1996) 101
\bi{afs}  P. C. Argyres, A. Faraggi and  A. Shapere, hep--th/9505190


\bi{bf1} F. Ferrari and A. Bilal, \np {\bf 469} (1996) 387





\bi{schw} W. Lerche, D.-J. Smit and N.P. Warner, \np {\bf 372} (1992) 87;\\
A. Klemm, B.H. Lian, S.S. Roan and S.T. Yau, hep--th/9407192

\bi{kklmv}  S. Kachru, A. Klemm, W. Lerche, P. Mayr and C. Vafa,
\np {\bf 459} (1996) 537

\bi{klmvw} A. Klemm, W. Lerche, P. Mayr, C. Vafa and N. Warner,  
hep--th/9604034

\bi{kv} S. Kachru and C. Vafa, \np{\bf 450} (1995) 69, hep-th/9505105

\bi{klm}  A. Klemm, W. Lerche and  P. Mayr, \pl {\bf 357} (1995) 313

\bi{al} P. Aspinwall and J. Louis, \pl {\bf 369} (1996) 233

\bi{ht} C.M. Hull and P.K. Townsend, \np {\bf 438} (1995) 109;\\
E. Witten, \np {\bf 443} (1995) 85


\bi{lw} W. Lerche and N.P. Warner,  hep--th/9608183

\bi{mw} E. Martinec and N. Warner, \np{\bf 459} (1996) 97, hep-th/9509161

\bi{klmw} W. Lerche,  P. Mayr and N. Warner, to appear

\bi{witten} E. Witten, hep--th/9507121


\bi{bsv} M. Bershadsky, C. Vafa and V. Sadov, hep--th/9510225

\bi{andy} A. Strominger, \np {\bf 451} (1995) 96


\bi{lll} K. Landsteiner, E. Lopez and D. A. Lowe, hep--th/9606146

\bi{porrati2} M. Porrati, hep--th/9607082, to appear in \pl
\bi{porrati} M. Porrati, \pl {\bf 377} (1996) 67;\\
G. Segal and A. Selby, \cmp {\bf 177} (1996) 775



\bi{f} F. Ferrari, hep--th/9609101

\bi{sen05} A. Sen, hep--th/9605150
\bi{ganor} O.J. Ganor, hep--th/9608109
\bi{dl} M. Douglas and M. Li, hep--th/9604041

\bi{bds} T. Banks, M. Douglas and N. Seiberg,  hep--th/9605199

\bi{sen08} A. Sen, hep--th/9608005

\bi{joh} A. Johansen, hep--th/9608186



\end{thebibliography}
\end{document}